\documentclass[twocolumn,a4paper,showpacs,preprintnumbers,amsmath,amssymb,nofootinbib,floatfix]{revtex4}
\input psfig.sty  
\def \be {\begin{equation}} 

\def \ee {\end{equation}} 
\def \bea {\begin{eqnarray}} 
\def \eea {\end{eqnarray}} 

\usepackage{graphicx}
\usepackage{dcolumn}
\usepackage{bm}
\usepackage{epsfig} 
\usepackage{color}

\begin{document}

\title{Galaxy cluster Sunyaev-Zel'dovich effect scaling-relation and type Ia supernova observations as a test for the cosmic distance duality relation}

\author{R. F. L. Holanda$^{1,2}$} \email{holandarfl@fisica.ufrn.br}
\author{L. R. Cola\c{c}o$^{1}$} \email{colacolrc@gmail.com}
\author{S. H. Pereira$^{3}$} \email{s.pereira@unesp.br}
\author{R. Silva$^{1,4}$} \email{raimundosilva@fisica.ufrn.br}

\affiliation{ $^1$Departamento de F\'{i}sica Te\'{o}rica e Experimental,\\ Universidade Federal do Rio Grande do Norte, 59300-000, Natal - RN, Brazil.\\
$^2$Departamento de F\'{i}sica, Universidade Federal de Sergipe, 49100-000, Aracaju - SE, Brazil,
\\$^3$Universidade Estadual Paulista (Unesp)\\Faculdade de Engenharia, Guaratinguet\'a \\ Departamento de F\'isica e Qu\'imica\\ Av. Dr. Ariberto Pereira da Cunha 333\\
12516-410 -- Guaratinguet\'a, SP, Brazil
\\$^4$Departamento de F\'{\i}sica, Universidade do Estado do Rio Grande do Norte, Mossor\'o, 59610-210, Brasil
}

\begin{abstract}

In this paper, we propose a new  test to the cosmic distance duality relation (CDDR), $D_L=D_A(1+z)^2$, where $D_L$ and $D_A$ are the luminosity and angular diameter distances, respectively. The data used correspond to 61 Type Ia Supernova luminosity distances and $Y_{SZE}-Y_X$ measurements of 61 galaxy clusters obtained by the {\it Planck} mission and the deep XMM-Newton X-ray data, where $Y_{SZE}$ is the integrated comptonization parameter  obtained via Sunyaev-Zel'dovich effect  observations and  $Y_X$ is the  X-ray counterpart. More precisely, we use the $Y_{SZE}D_{A}^{2}/C_{XSZE}Y_X$ scaling-relation and a deformed CDDR, such as $D_L/D_A(1+z)^2=\eta(z)$, to verify if $\eta(z)$ is compatible with the unity. Two $\eta(z)$ functions are used, namely, $\eta(z)=1+\eta_0 z$ and $\eta(z)=1+\eta_0 z /(1+z)$. { We obtain that the CDDR validity ($\eta_0=0$)  is verified within $\approx 1.5\sigma$ c.l. for both $\eta(z)$ functions.}

\end{abstract}
\pacs{98.80.-k, 95.36.+x, 98.80.Es}
\maketitle

\section{Introduction}

The concept of distance in cosmology is of fundamental importance when one wants to relate observational data with theoretical models. In particular, two types of distance are of great importance in observational cosmology, namely, the luminosity distance, $D_L$, and the angular diameter distance,  $D_A$. The first is a distance measurement of an object with basis on the decrease of its brightness with the distance and the second one is related with the measure of the angular size of the object projected on the celestial sphere. Both distances depend on the redshift $z$ of the object studied and they are related by:
\begin{equation}
{D_L(z)\over D_A(z)(1+z)^2}=1 \label{DLDA}.
\end{equation}
This result is known as cosmic distance duality relation (CDDR), which is a version of Etherington’s reciprocity law in the context of astronomical observations \cite{ETHE}. Such relation is easily obtained in a Friedmann-Robertson-Walker (FRW) background, nevertheless it is completely general, valid for all cosmological models based on Riemannian geometry, requiring solely observer and source be connected just by null geodesics and that the number of photons is conserved over the cosmic evolution \cite{BASSET}. Such generality places this relationship as being of fundamental importance in observational cosmology and any deviation from it may indicate the possibility of a new physics or the presence of systematic errors in observations \cite{ELLIS}.

Along with  unprecedented increase in number and quality of astronomical data,  different methods have been proposed  in order to test the validity of the CDDR.  One may roughly divide them in two classes, cosmological model-dependent tests \cite{bernardis,uzan2005,avgoustidis2010,avgousti2012,hol2011,more2016,piazza2016}, usually performed within the $\Lambda$CDM model, and cosmological model-independent ones. The observations of type Ia Supernovae (SNe Ia), cosmic background radiation (CMB), baryon acoustic oscillations (BAO), galaxy cluster gas mass fraction, angular diameter distance of galaxy clusters, strong gravitational lensing, compact radio sources and $H(z)$ data jointly with the parametrization ${D_L(z)\over D_A(z)(1+z)^2}=\eta(z)$ have been explored in literature in order to perform  cosmological model-independent tests \cite{hol2010,PUXUN,liwu2011,gon2012,hol20121,hol20122,costa20151,chen,hol20161,jailson2011,meng2012,yang2013,ELLIS2013,jhingan,LIOO,holper2017,yizheng2014,holbarros2016,shafieloo2013,rana,rana20162,linli2018,hol20172}. The most used $\eta(z)$ functions have been:
\begin{itemize}
\item (i) ~~ $\eta(z)=1+\eta_0 z$

\item (ii) ~~ $\eta(z)=1+\eta_0 {z \over (1+z)}$.
\end{itemize}
The first one is a continuous and smooth one-parameter linear expansion, whereas the second one includes a possible epoch-dependent correction, which avoids the divergence at very high redshifts. These two parametrizations recover the standard case for $\eta_0 = 0$ \cite{hol2011,jhingan}.

As commented earlier, several authors have proposed different approaches to test the CDDR validity by using galaxy clusters observations. For instance, Ref.\cite{hol2010} used angular diameter distance samples of  galaxy clusters obtained via their Sunyaev-Zel'dovich effect (SZE) and X-ray observations jointly with luminosity distances  of SNe Ia and proposed a  cosmological  model-independent test for the CDDR. As result, it was showed that  the isothermal ellipsoidal model is a better geometrical hypothesis describing the structure of galaxy clusters compared with the spherical model if CDDR is valid (see also Ref.\cite{PUXUN,liwu2011}). The Ref.\cite{gon2012} showed that the  gas mass fraction of galaxy clusters obtained from their X-ray observations also depends on the CDDR validity and proposed a test involving this kind of measurement and SNe Ia observations (see also Ref.\cite{hol20121}). In order to avoid two different types of astronomical observations, the Ref.\cite{hol20122} proposed a test that uses exclusively gas mass fraction measurements of galaxy clusters obtained via SZE and X-ray observations. Applying gaussian process, the Ref.\cite{costa20151} proposed a test based on galaxy clusters observations and $H(z)$ measurements (see also \cite{chen}). Up to now no significant departure from the CDDR validity was verified (a table with  recent estimates from different methods and observations can be found in Ref.\cite{hol20161}), however, the point number in galaxy cluster samples used are lower than 40 points and the limits on $\eta_0$ arising from such observations are not so restrictive. Then, new methods with different astronomical observations and redshift range are still welcome to validate the whole cosmological framework as well as to search  systematic errors in astronomical data.

In this paper, we propose a new test to the CDDR by using galaxy cluster SZE scaling-relation and SNe Ia observations. Scaling-relations in galaxy clusters result from the hierarchical structure formation theory  when gravity is the dominant process \cite{KAISER1986}. Particularly,  we consider the following scaling-relation\footnote{{As commented earlier, the Ref.\cite{hol2010} used angular diameter distance samples of  galaxy clusters obtained via their Sunyaev-Zel'dovich effect (SZE) and X-ray observations to test the CDDR. However, in that case, the $D_A$ for each galaxy cluster is observationally known. In our work, the constant $C$ is not observational and the $D_A$ quantity for each cluster can not be directly obtained.}}: $Y_{SZE}D_{A}^{2}/C_{XZS}Y_X = C $, where $Y_{SZE}D_{A}^{2}$ is the integrated comptonization parameter of a galaxy cluster obtained via SZE observations multiplied by its angular diameter distance,  $Y_X$ is the  X-ray counterpart and $C_{XSZE}$ is a constant, with $C$ an arbitrary constant \cite{KAISER1986,GALLI2013}. In a very recent paper, the authors of the Ref.\cite{hola2019} derived a new expression for this ratio when there is a possible departure from the CDDR validity and/or a variation of the fine structure constant $\alpha$, with $Y_{SZE}D_{A}^{2}/C_{XSZE}Y_X = {\rm{C}} \alpha^{3.5} \eta^{-1}(z)$, where $\eta(z) = D_L/D_A(1+z)^2$. On the other hand, if one considers the class of theories with a non minimal multiplicative coupling between the usual electromagnetic part of matter fields and a new scalar field it is possible to obtain $\alpha(z) \propto  \eta(z)^2$  (these theories explicitly break the Einstein equivalence principle in the electromagnetic sector, see next section for details). Thus, one may show that  $Y_{SZE}D_{A}^{2}/C_{XSZE}Y_X = {\rm{C}} \eta^{6}(z)$. In this way, by using  $Y_{SZE}-Y_X$ measurements of 61 galaxy clusters  taken from the Ref.\cite{ADE2011}, 61 SNe Ia luminosity distances  taken from the Ref.\cite{SNE} in the galaxy cluster redshifts and the relation ${D_L(z)\over D_A(z)(1+z)^2}=\eta(z)$,  we put limits on the $\eta_0$ parameter for the most used $\eta(z)$ functions,  as (i) and (ii). {For both $\eta(z)$ functions is obtained  $\eta_0=0$ within 1.5$\sigma$ c.l..}

The paper is organized as follows. In Section 2 we present the methodology, Section 3 contains the data used in our analyses. Section 4 presents the analyses and results, and the conclusions are given in Section 5.

\section{Methodology}

In order to obtain  the key relation used in our test, let us discuss briefly the method presented in the Ref.\cite{hola2019}. 

The scaling-relations in galaxy clusters rise from the simplest model for formation of structures, when gravity is the dominant process. In this scenario,  simple scaling-relations between basic galaxy cluster properties and the total mass are predicted by self-similar models (details can be found in the Ref.\cite{KAISER1986}). In our work, we are interested in the scaling-relation involving the SZE and X-ray surface brightness \cite{sze1,birk}  $Y_{SZE}D_{A}^{2}/C_{XZS}Y_X = C $.

The SZE is a distortion caused in the CMB spectrum  and is proportional to the Compton parameter $y$, which quantifies the gas pressure of the intracluster medium integrated along the line of sight \cite{sze1,birk,carls}. By integrating it over the solid angle of a galaxy cluster $(d\Omega = dA/D_{A}^2$), it is possible to obtain the integrated Compton parameter $Y_{SZE}$, such as:
\begin{equation} \label{ysz}
Y_{SZE} \equiv \int_{\Omega} yd\Omega,
\end{equation}
or, equivalently,
\begin{equation} \label{ysz}
Y_{SZE}D^2_A \equiv \frac{\sigma_T}{m_ec^2} \int P dV,
\end{equation}
where $P=n_eK_BT$ is the integrated thermal pressure of the intracluster gas along the line of sight. Then, as commented in the Ref.\cite{GALLI2013}, one may see  that $Y_{SZE} D_{A}^{2}$ has a dependency on the fine structure constant, $\alpha$, through the Thompson cross section as:
\begin{equation} \label{ysz2}
Y_{SZE}D_{A}^{2} \propto \alpha^2.
\end{equation}

\begin{figure*}
\includegraphics[width=0.45\textwidth]{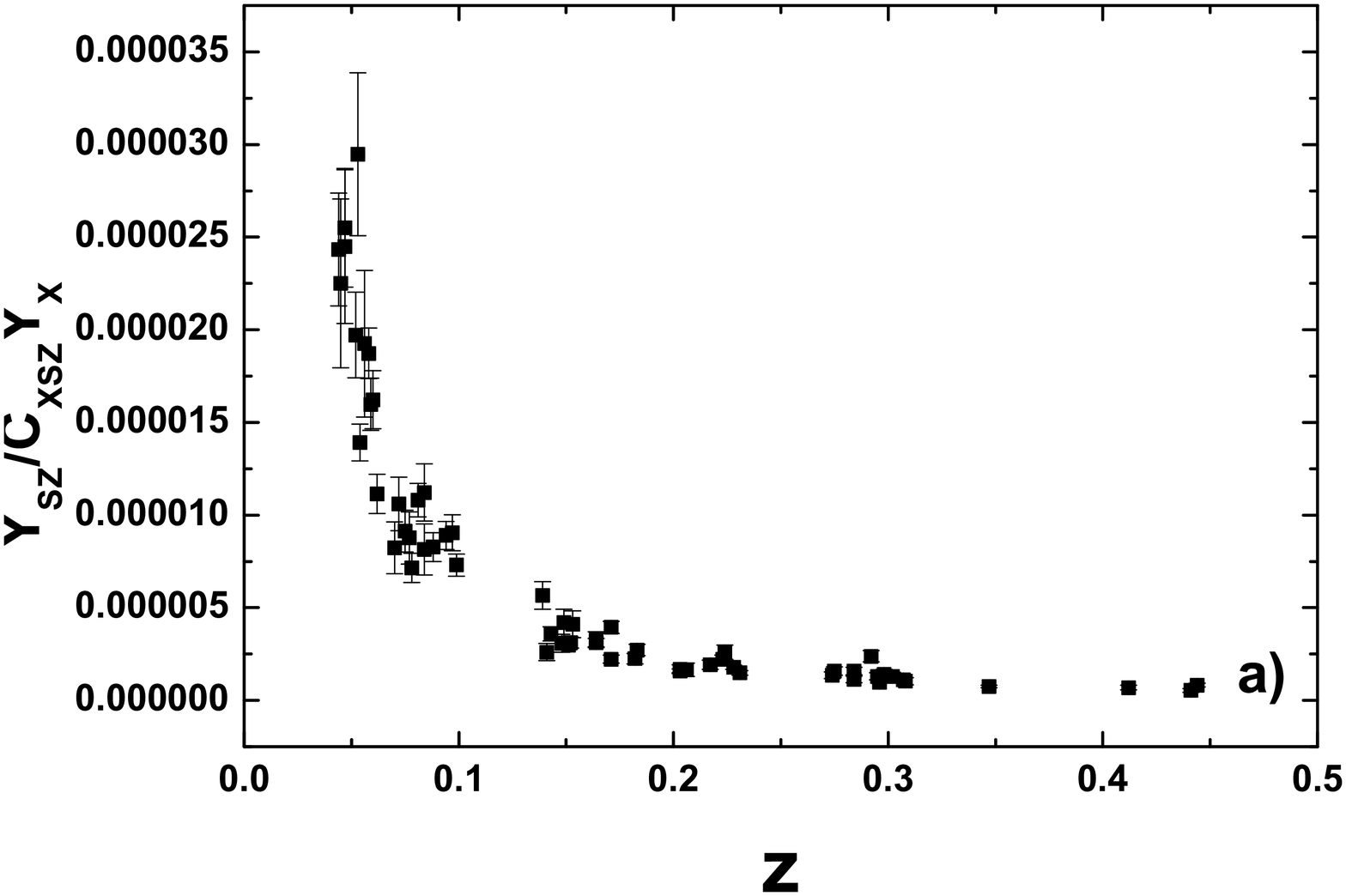}
\includegraphics[width=0.45\textwidth]{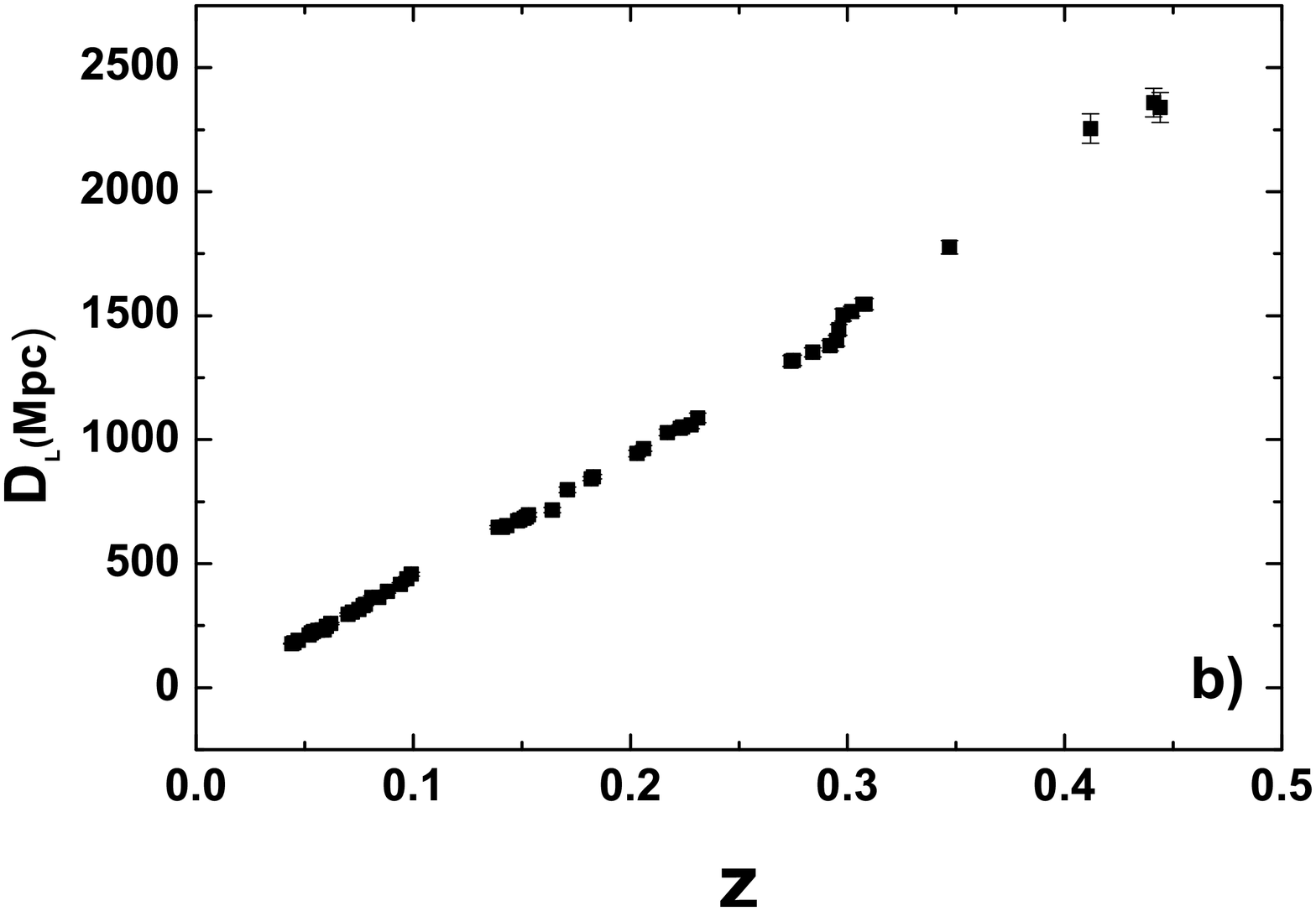}
\caption{The Figures (a) and (b) show the $Y_{SZE}/C_{XSZE}Y_X$ and SNe Ia data used in our analyses, respectively.}
\end{figure*}

On the other hand, the $Y_X$ parameter, obtained through X-ray surface brightness observations, is defined as: 
\begin{equation}
\label{yx}
Y_X = M_g(R)T_X,
\end{equation}
where $T_X$ is the spectroscopically determined X-ray temperature  and $M_g(R)=\mu_em_p\int n_e dV$  is the gas mass within the radius $R$, in this expression $m_p$ stands for the proton mass and  $\mu_e$ corresponds to the mean molecular weight of electrons. The key quantity here is $M_g(R)$, which can be written in terms of the fine structure constant and the CDDR  as (see e.g. \cite{GALLI2013,hola2019} for details):
\begin{equation}
M_g(<R) \propto \alpha(z)^{-3/2} D_LD_{A}^{3/2}.
\end{equation}
Thus, if one considers any departure from the CDDR, for instance,  $D_L/(1+z)^2D_A=\eta(z)$,  $M_g(R)$ and, consequently, $Y_X$, will depend on $\alpha(z)$ and $\eta(z)$ as:
\begin{equation}
Y_X \propto M_g(<R) \propto \alpha(z)^{-3/2} \eta(z).
\end{equation}
As commented earlier, the authors of Refs.\cite{hees2,minazzoli,hees3}  considered a wide class of theories of gravity that explicitly breaks the Einstein equivalence principle in the electromagnetic sector. They consider models which implement the break of the equivalence principle by introducing an additional term into the action, coupling the usual matter fields  to a new scalar field, which is motivated by scalar-tensor theories of gravity. In this context,  the entire electromagnetic sector is affected and $\alpha(z)$ and $\eta(z)$ are intimately and unequivocally linked by (see also \cite{brans,damour,damour3,fujii,laur}):
\begin{equation}
\frac{\Delta \alpha}{\alpha}(z) \equiv \frac{\alpha (z)-\alpha_0}{\alpha_0}= \eta(z)^2 -1.
\end{equation}
Then, the equations (5) and (8) depend on $\eta(z)$  as:
\begin{eqnarray} \label{YXYSZ}
Y_{SZE}D_{A}^{2} (z) &\propto & \eta(z)^{4} \\
Y_X &\propto & \eta(z)^{-2}.
\end{eqnarray}

Our method is based on the $Y_{SZE}D_{A}^{2}/C_{XSZE}Y_X=C$ scaling-relation\footnote{In this expression: $C_{XSZE} = \frac{\sigma_T}{m_ec^2} \frac{1}{\mu_em_p} \approx 1.416 .10^{-19} \frac{{\rm{Mpc}}^2}{M_{\odot}{\rm{keV}}}$.}. As it is largely known,  $Y_{SZE}$ and $Y_X$ are approximations of the thermal energy of the cluster. This ratio is expected to be  constant with redshift since $Y_{SZE}D^2$ and $Y_X$ are expected to scale in the same way with mass and redshift as power-laws \cite{Kravtsov,Stanek,Fabjan,Kay,Bohringer,Nagai}. Moreover, if galaxy clusters are isothermal this ratio would be exactly equal to unity, or constant with redshift if the galaxy clusters have a universal temperature profile. Actually, numerical simulations have shown that this ratio has small scatter, at the level of $\approx 15\%$ \cite{Stanek,Fabjan,Kay}.  Then, as one may see, this ratio may be  written as:
\begin{equation}
\frac{Y_{SZE}D_{A}^{2}}{C_{XSZE}Y_X} = C\eta(z)^6.
\end{equation} 

{In this point, it is worth to stress that the $Y_X$ quantity is proportional to $ M_g$, which depends on the galaxy cluster  distance such as: $M_g \propto D_LD_{A}^{3/2}$. Usually, this measurement is obtained by using a fidutial ($F$)  flat $\Lambda$CDM model with $\Omega_M=0.3$ and $H_0=70$ km/s/Mpc, and the CDDR validity, resulting in: $M_g \propto D_{AF}^{5/2}$. Then, in order to eliminate the dependence of $M_g$ with respect to the fidutial model, we multiply the $Y_X$ quantity by $D_{A}^{5/2}/D_{AF}^{5/2}$ and the Eq.(11) becomes:

\begin{equation}
\frac{Y_{SZE}D_{AF}^{5/2}}{D_{A}^{1/2}C_{XSZE}Y_X} = C\eta(z)^6.
\end{equation}
Finally, to perform our test and to impose limits on  $\eta(z)$ functions it is necessary to know  $D_A$ for each galaxy cluster in the sample. This quantity is obtained by using SNe Ia luminosity distances with identical redshifts to those of the clusters and  considering a deformed CDDR, such as $D_A=\eta(z)^{-1}(1+z)^{-2}D_L$. Then, one may obtain:

\begin{equation}
\frac{Y_{SZE}D_{AF}^{5/2}(1+z)}{D_{L}^{1/2}C_{XSZE}Y_X} = C\eta(z)^{5.5}.
\end{equation} 
This is our key equation.} Since one knows the luminosity distances of  galaxy clusters from a sample with their $Y_{SZE}-Y_X$ quantities measured, it is possible to use  this expression to impose limits on the $\eta(z)$ functions.  We obtain limits on the $\eta_0$ parameter  for the two $\eta(z)$ functions, namely (i) and (ii).  We also put limits on the constant $C$ in order to obtain some information about the thermal energy of the cluster.

\begin{figure*}[htb]
\includegraphics[width=0.3\textwidth]{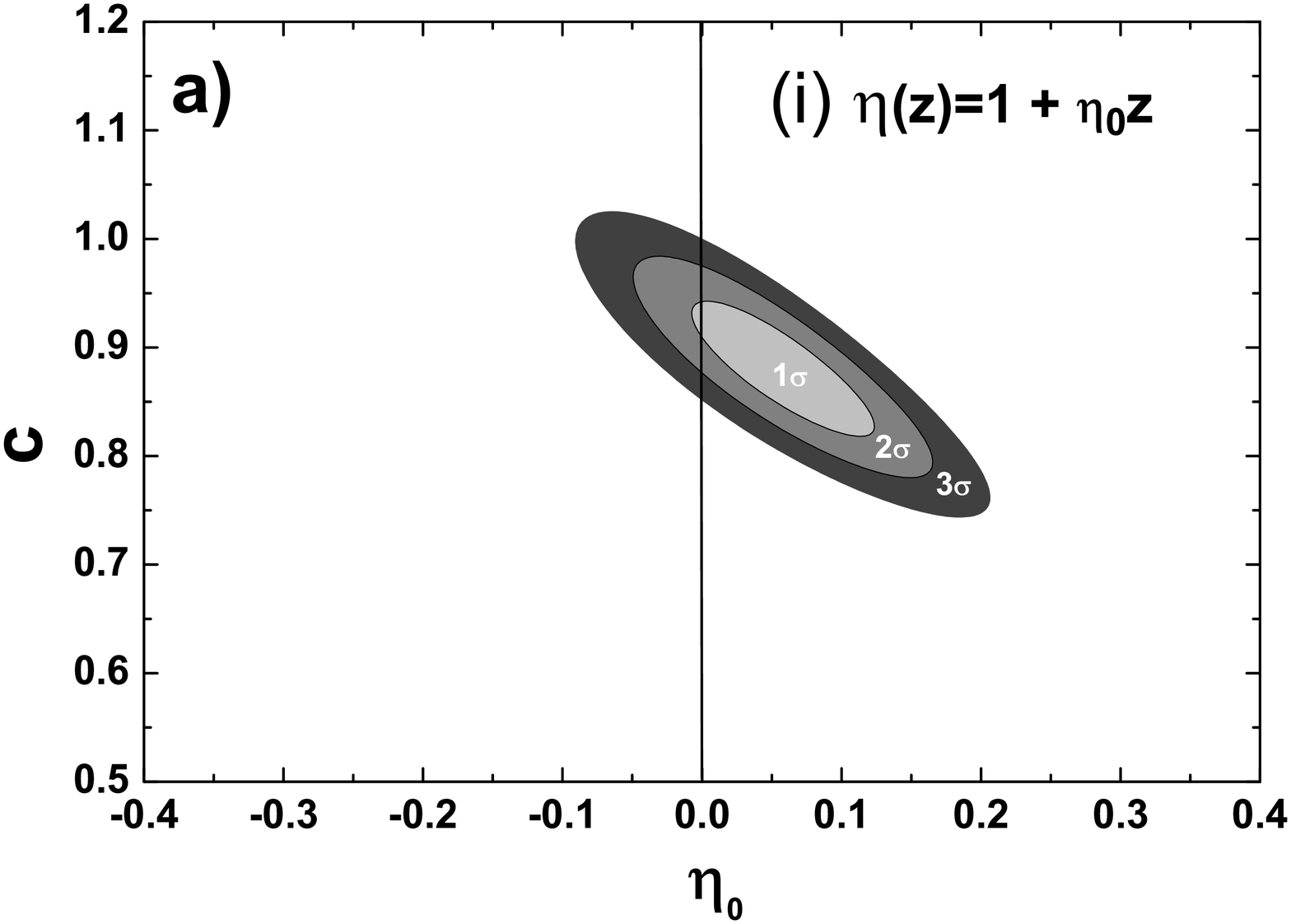}
\includegraphics[width=0.3\textwidth]{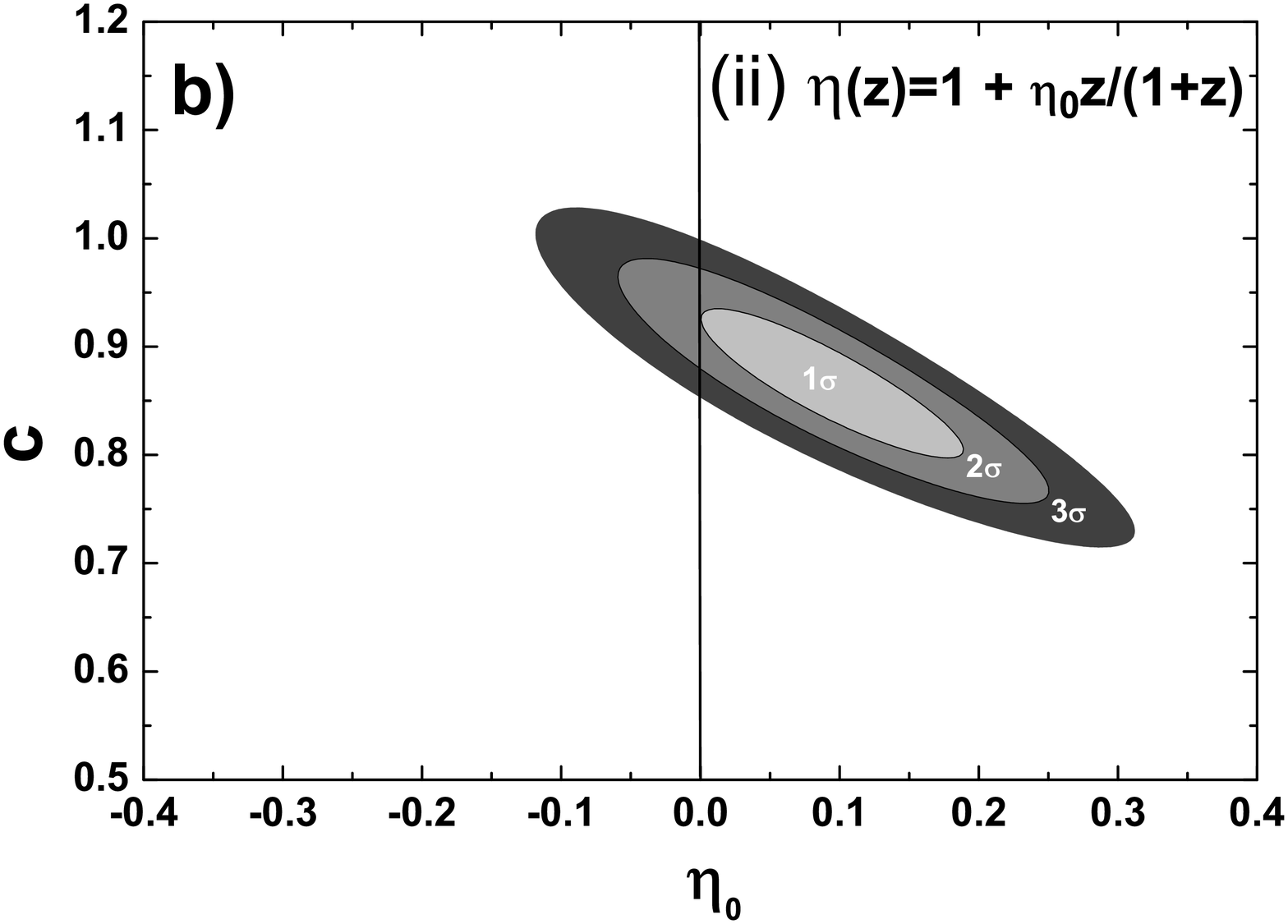}
\includegraphics[width=0.3\textwidth]{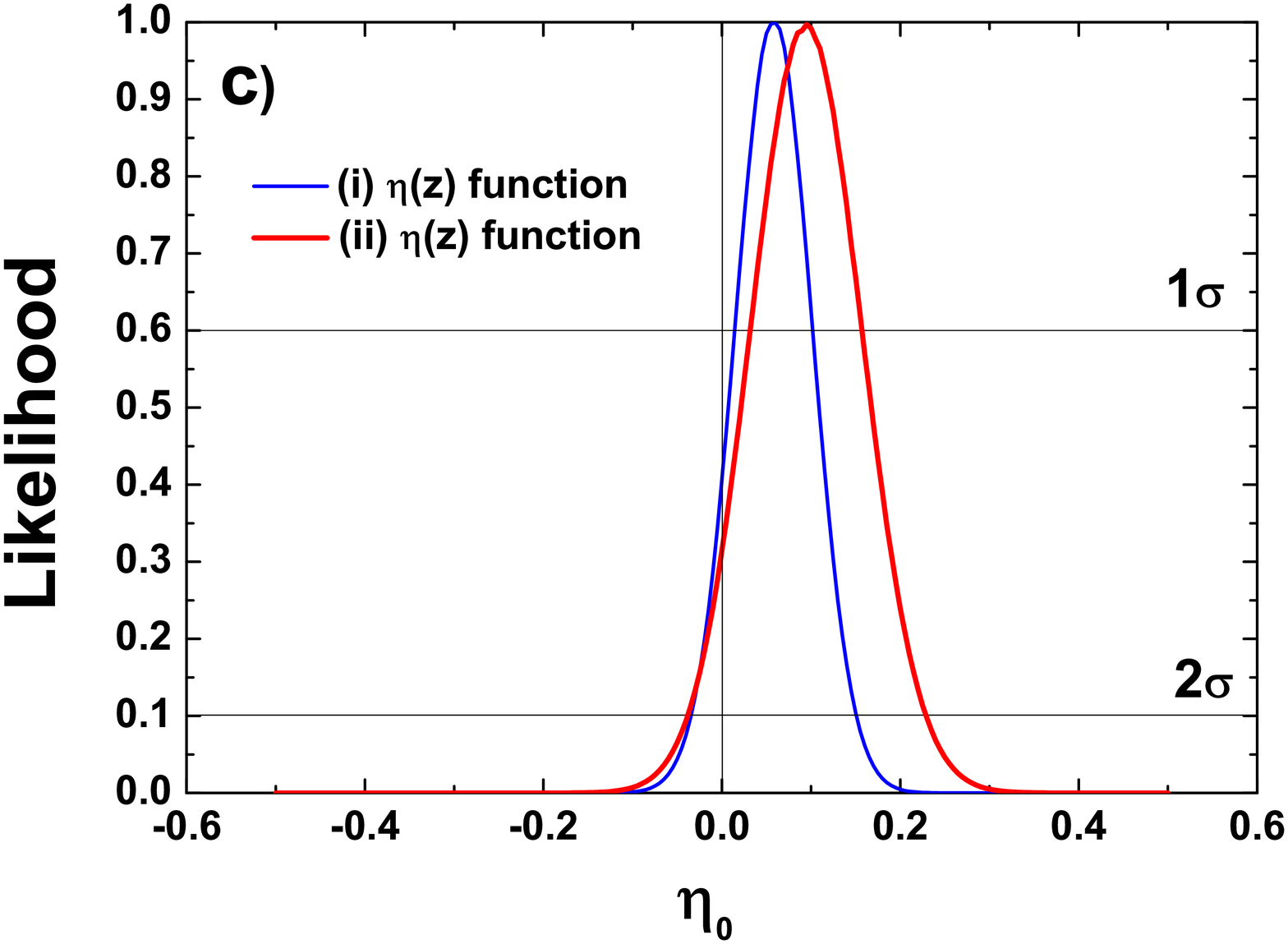}
\caption{  The Figures (a) and (b) show the 1$\sigma$, 2$\sigma$ and 3$\sigma$ c.l. regions on the $(C,\eta_0)$ plane for each $\eta(z)$ function. The Fig. (c) shows the likelihood for $\eta_0$ (by marginalizing on $C$).}
\end{figure*}

\section{Samples}

Our CDDR test is performed by using the following samples:

\begin{itemize}
\item Galaxy clusters: we use $Y_{SZE}-{Y_X}$  measurements of 61 galaxy clusters obtained from the first {\it Planck mission} all-sky data set jointly with deep XMM-Newton archive observations (see Fig.1a). The galaxy clusters were detected at high  signal-to-noise  within the following redshift interval and mass, respectively: $0.044 \leq z \leq 0.444$ and $2 \mbox{x} 10^{14} M_{\odot} \leq M_{500} \leq 2 \mbox{x} 10^{15} M_{\odot}$, where $M_{500}$ is the total mass corresponding to a total density contrast of $500\rho_c(z)$, where $\rho_c(z)$ is the critical density of the Universe at the cluster redshift. Actually, the quantities $Y_X$ and $Y_{SZE}$ were determined within the $R_{500}$ (at which the mean enclosed mass density is equal to 500 cosmological critical density). In order to estimate the $Y_{SZE}-{Y_X}$  measurements  of a galaxy cluster, one needs to add some complementary assumptions about their physical properties. Many studies about the intracluster gas and dark matter distribution in galaxy clusters have been performed \cite{arnaud,giula,jeniffer}. Assigning a single temperature to the whole cluster is a gross approximation \cite{viki}. The importance of the intrinsic geometry of the cluster, for instance, has been emphasized by many authors \cite{morandi,limousin,holandaap}, in particular, non-sphericity results in bias in mass estimates. { Then, it is important to stress that  a violation of CDDR in this type of analysis would not necessarily indicate new fundamental physics but more likely would point to the failure of cluster models/assumptions and need for better modeling.}

The thermal pressure ($P$) of the intracluster medium for each galaxy cluster used in our analyses was modeled by the Ref.\cite{ADE2011} via the universal pressure profile discussed in the Ref.\cite{arnaud}. This universal profile was obtained by comparing  observational data (a representative sample of nearby clusters covering the mass range $10^{14}M_{\odot} < M_{500} < 10^{15} M_{\odot}$) with simulated data. The $T_X$ quantity was measured in the $[0.15-0.75] R_{500}$ region. By using $D_A$ calculated from the {\it Planck mission} flat $\Lambda$CDM framework, the Ref.\cite{ADE2011} showed that the $\frac{Y_{SZE}D_{A}^{2}}{C_{XSZE}Y_X}$ ratio for the galaxy clusters considered in this work has very small scatter, at the level of $\approx 15\%$. Moreover, it was also verified that this scaling-relation does not seem to depend crucially on the dynamical state of the clusters. 
\item SNe Ia: we use a sub-sample of the latest and largest Pantheon Type Ia supernovae sample in order to obtain $D_L$ of the galaxy clusters. The Pantheon SNe Ia compilation consist of  1049  spectroscopically confirmed SNe Ia covering the redshift
range $0.01 \leq z \leq 2.3$ \cite{SNE}. To perform our test, we need to use SNe Ia and galaxy clusters in the identical redshifts. Thus, for each galaxy cluster, we select SNe Ia with redshifts obeying the criteria $|z_{GC} - z_{SNe}| \leq 0.005$ and calculate the following weighted average for the SNe Ia data:
\begin{equation}
\begin{array}{l}
\bar{\mu}=\frac{\sum\left(\mu_{i}/\sigma^2_{\mu_{i}}\right)}{\sum1/\sigma^2_{\mu_{i}}} ,\hspace{0.5cm}
\sigma^2_{\bar{\mu}}=\frac{1}{\sum1/\sigma^2_{\mu_{i}}}.
\end{array}\label{eq:dlsigdl}
\end{equation}
\end{itemize}
We ended with 61 measurements of $\bar{\mu}$ and $\sigma^2_{\bar{\mu}}$. The luminosity distance for each galaxy cluster  is obtained through $D_L (z)=10^{(\bar{\mu} (z)-25)/5}$ and $\sigma_{D_L}^{2}= \left( \frac{\partial D_L}{\partial \bar{\mu}} \right)^2\sigma^2_{\bar{\mu}}$ is the associated error over $D_L$  (see Fig.1b). 

\begin{table*}[ht]
\caption{A summary of the current constraints on the $\eta_0$ parameter from different methods by using galaxy cluster observations and the (i) and (ii) $\eta(z)$ functions. The ADD and GMF correspond to angular diameter distance and gas mass fraction, respectively. The * and ** symbols correspond to results with 1$\sigma$ and 2$\sigma$ c.l., respectively. }
\label{tables1}
\par
\begin{center}
\begin{tabular}{|c||c|c|c|c|}
\hline\hline  Reference &  Data Sample & (i) function &   (ii) function 
\\ \hline\hline 
Holanda et al. (2010)** & ADD + SNe Ia & $-0.28^{+0.44}_{-0.44}$ & $-0.43^{+0.60}_{-0.60}$\\
Xiang et al. (2011)* & ADD + SNe Ia & $-0.15^{+0.17}_{-0.17}$ & $-0.23^{+0.24}_{-0.24}$\\
Puxun et al. (2011)* & ADD + SNe Ia & $-0.07^{+0.19}_{-0.19}$ & $-0.11^{+0.26}_{-0.26}$\\
Gon\c{c}alves et al. (2012) & GMF + SNe Ia  &$-0.03^{+1.03}_{-0.65}$& $-0.08^{+2.28}_{-1.22}$ \\
Holanda et al. (2012) & GMF & $-0.06 \pm 0.16$ & $-0.07 \pm 0.24$  \\
Liang et al. (2013)** & ADD + SNe Ia & $-0.232 \pm 0.232$ & $-0.351\pm{0.368}$\\
Yang et al. (2013) & ADD   + SNe Ia      & $0.16^{+0.56}_{-0.39}$    & - \\
S.-da-Costa et al. (2015) & ADD  + $H(z)$   & $-0.100^{+0.117}_{-0.126}$ & $-0.157^{+0.179}_{-0.192}$  \\
S.-da-Costa et al. (2015) & GMF  + $H(z)$ & $0.062^{+0.168}_{-0.146}$ & $-0.166^{+0.337}_{-0.278}$   \\
Chen et al. (2015)& ADD + SNe Ia + $H(z)$ &$0.07 \pm 0.08$ & $0.15 \pm 0.18$   \\
Holanda \& Pereira (2016)*& ADD + SNe Ia + $H(z)$ &$0.07 \pm 0.106$ & $0.097 \pm 0.152$   \\
This paper** & $Y_{SZE}-Y_X$ + SNe Ia & $\eta_0=0.05 \pm 0.07$ & $0.09 \pm 0.16$ \\
\hline\hline
\end{tabular}
\end{center}
\end{table*}

\section{analyses and results}

We evaluate our statistical analysis by defining the likelihood distribution function, ${\cal{L}} \propto e^{-\chi^{2}/2}$, 

\begin{equation}
\chi^2 = \sum_{i=1}^{61} \frac{\left[ C\eta(z)^{5.5} - \frac{D_{AF}^{5/2}Y_{SZE}(1+z)}{C_{XSZE}Y_XD_L^{1/2}}\right]^2}{\sigma_{i,obs}^{2}},
\end{equation}
where $\sigma_{i,obs}$  stands for the statistical errors of $Y_{SZE}$, $Y_X$ and $D_L$. We added in quadrature to statistical error of the $\frac{Y_{SZE}}{C_{XSZE}Y_X}$ quantity   a $15\%$ error attributed to an additional intrinsic scatter in order to obtain the $\chi^2_{red}\approx 1$. Two cases are considered for $\eta(z)$, namely: (i) $\eta(z)=1+\eta_0z$ and (ii) $\eta(z)=1+\eta_0z/(1+z)$.

 Our results are plotted in Fig.(2). The Figures (2a) and (2b) show the 1$\sigma$, 2$\sigma$ and 3$\sigma$ c.l. regions on the $(C,\eta_0)$ plane for each $\eta(z)$ function. From the Fig.(2a), we obtain at 1$\sigma$, 2$\sigma$ and 3$\sigma$ c.l. (two free parameters): $\eta_0=0.05 \pm 0.055 \pm 0.10 \pm 0.16$ and  $C=0.86 \pm 0.07 \pm 0.13 \pm 0.17 $ with $\chi^2=66.12$. From the Fig.(2b),  we obtain at 1$\sigma$, 2$\sigma$ and 3$\sigma$ c.l. (two free parameters): $\eta_0=0.09 \pm 0.09 \pm 0.16 \pm 0.25$ and  $C=0.85 \pm 0.08 \pm 0.14 \pm 0.18$ with $\chi^2=65.17$. As one may see,  our estimates on the $C$ parameter are  only marginally in agreement with an isothermal assumption for the galaxy clusters, which correspond to $C=1$. Moreover, the  $\eta_0 = 0$ value is allowed  within $1\sigma$ c.l..

Fig.(2c) shows the likelihood for $\eta_0$ (by marginalizing on $C$). We obtain for the (i) and (ii) $\eta(z)$ functions, respectively, at 1$\sigma$ and 2$\sigma$ c.l.: $\eta_0 = 0.05 \pm 0.04 \pm 0.07 $ (blue line) and  $\eta_0 = 0.09 \pm 0.06 \pm 0.013$ (red line). Then,  the CDDR validity ($\eta_0=0$)  is obtained within $1.5\sigma$, which indicates no  tension between the data if the CDDR is taken as valid.

Some clusters exhibit the so-called cool cores, central regions of very dense gas where the cooling time is less than the Hubble time. By checking the Table I in the Ref.\cite{ADE2011} we find that 22  and 39 galaxy clusters, respectively, are cool and non-cool cores. Then, we also perform our analyses by using these two sub-groups separately. Naturally, the error bars from the new results are larger than those by using the complete sample. From the cool core sub-sample, we obtain for (i) and (ii) $\eta(z)$ functions, respectively: $\eta_0 = 0.045 \pm 0.070 \pm 0.095$ and  $\eta_0 = 0.08 \pm 0.10\pm 0.15$ at 1$\sigma$ and 2$\sigma$ c.l.. From the non-cool core sub-sample, we obtain for linear and non-linear $\eta(z)$ functions, respectively: $\eta_0 = 0.065 \pm 0.070 \pm 0.10$ and  $\eta_0 = 0.09 \pm 0.011 \pm 0.13$. Then, the results are in full agreement each other and the CDDR validity  is verified.

\section{Conclusions}

In this work, we presented a cosmological model-independent test to the cosmic distance duality relation (CDDR) by using galaxy cluster Sunyaev-Zel'dovich scaling-relation data jointly with SNe Ia observations. The galaxy cluster sample consisted of 61  $Y_{SZE}-Y_X$  measurements  obtained from the first {\it Planck} all-sky data set jointly with deep XMM-Newton archive observations within the redshift interval $0.044 \leq z \leq 0.444$.  By using a deformed CDDR such as  $D_L/D_A(1+z)^2=\eta(z)$, we showed that it is possible to test the CDDR with the galaxy cluster Sunyaev-Zel'dovich effect scaling-relation and SNe Ia via the equation $\frac{Y_{SZE}D_{AF}^{5/2}(1+z)}{D_{L}^{1/2}C_{XSZE}Y_X} = C\eta(z)^{5.5}$. The luminosity distance for each galaxy cluster was obtained by using a sub-sample of the latest and largest Pantheon SNe Ia sample. 

For $\eta(z)$ we used two functions widely considered in the literature, namely, $\eta(z)=1+\eta_0 z$ and $\eta(z)=1+\eta_0 z /(1+z)$ and put observational limits on $\eta_0$. Moreover,  limits on the constant $C$ were also performed. For both $\eta(z)$ functions it was obtained that $\eta_0=0$ is verified within  $1.5\sigma$ c.l. (marginalizing on $C$). On the other hand,  $C=1$ is only marginally compatible with the data, indicating a departure from an isothermal assumption  for the temperature profile of the galaxy clusters used in our analysis. Since  some clusters exhibit  cool or non-cool cores, we also performed our analyses by using these two sub-groups separately and the results obtained for $\eta_0$ are in full agreement each other without evidence for a CDDR violation. { Then, our results did not depend  on the dynamical state of the clusters at current level of accuracy}. 

\acknowledgements

RFLH thanks financial support from  {Conselho Nacional de Desenvolvimento Cient\'ifico e Tecnol\'ogico} (CNPq) (No. 428755/2018-6 and 305930/2017-6). SHP thanks  financial support from CNPq (No. 303583/2018-5 and 400924/2016-1) and CAPES. RS thanks CNPq (Grant No. 303613/2015-7) for financial support.

\label{lastpage}
\end{document}